\documentclass[conference, anonymous]{IEEEtran}

\usepackage[letterpaper,top=2cm,bottom=2cm,left=3cm,right=3cm,marginparwidth=1.75cm]{geometry}

\newcommand{\sectopic}[1]{\vspace{0.2em}\par\noindent{\textit{\bfseries #1}}}

\usepackage[utf8]{inputenc}
\usepackage{array}
\usepackage{booktabs}
\usepackage{multicol}
\usepackage{multirow}
\usepackage{listings}
\usepackage{tabularx}
\usepackage{longtable}
\usepackage{threeparttable}
\usepackage{enumitem} 
\usepackage{tcolorbox}
\usepackage{balance}
\usepackage{hyperref}
\usepackage[utf8]{inputenc}
\lstdefinestyle{prompt}{
    basicstyle=\ttfamily\footnotesize,
    backgroundcolor=\color{gray!10},
    frame=single,
    breaklines=true,
    keepspaces=true,
    columns=fullflexible,
    captionpos=b
}

\begin{document}
\pagenumbering{arabic}
%
\title{
From Domain Documents to Requirements: Retrieval-Augmented Generation in the Space Industry}


\author{
    \IEEEauthorblockN{Chetan Arora\IEEEauthorrefmark{1}, Fanyu Wang\IEEEauthorrefmark{1}, 
    Chakkrit  Tantithamthavorn\IEEEauthorrefmark{1},      
    Aldeida Aleti\IEEEauthorrefmark{1},
    Shaun Kenyon\IEEEauthorrefmark{2}
  }
    \IEEEauthorblockA{\IEEEauthorrefmark{1}Monash University, Melbourne, Australia}
    \IEEEauthorblockA{\IEEEauthorrefmark{2}Starbound Space Solutions, Queensland, Australia}
    Email: 
    \{chetan.arora,
    fanyu.wang,
    chakkrit,
    aldeida.aleti\}@monash.edu,
    shaun@starboundsolutions.com
\vspace*{-1em}
}


%


\maketitle

\thispagestyle{plain}
\pagestyle{plain}

\begingroup\renewcommand\thefootnote{\textsection}
\endgroup

\begin{abstract}
Requirements engineering (RE) in the space industry is inherently complex, demanding high precision, alignment with rigorous standards, and adaptability to mission-specific constraints. Smaller space organisations and new entrants often struggle to derive actionable requirements from extensive, unstructured documents such as mission briefs, interface specifications, and regulatory standards. 
In this innovation opportunity paper, we explore the potential of Retrieval-Augmented Generation (RAG) models to support and (semi-)automate requirements generation in the space domain. We present a modular, AI-driven approach that preprocesses raw space mission documents, classifies them into semantically meaningful categories, retrieves contextually relevant content from domain standards, and synthesises draft requirements using large language models (LLMs). We apply the approach to a real-world mission document from the space domain to demonstrate feasibility and assess early outcomes in collaboration with our industry partner, Starbound Space Solutions.
Our preliminary results indicate that the approach can reduce manual effort, improve coverage of relevant requirements, and support lightweight compliance alignment. We outline a roadmap toward broader integration of AI in RE workflows, intending to lower barriers for smaller organisations to participate in large-scale, safety-critical missions.

\end{abstract}


%
\IEEEpeerreviewmaketitle

\section{Introduction}~\label{sec:introduction}
\vspace*{-0.1em}
Modern systems engineering projects in safety-critical domains—such as aerospace, defence, and satellite communications—involve increasingly complex requirements documentation and compliance landscapes. Requirements engineering (RE) plays a pivotal role in ensuring safety, reliability, and mission success~\cite{hirshorn2017nasa}. However, the increasing complexity and scale of documentation in these domains create a significant bottleneck in how requirements are authored, analysed, and aligned with operational scenarios and domain and regulatory standards. These documents—often distributed across hundreds of pages in semi-structured or unstructured formats—are difficult to process using existing RE or general-purpose natural language processing (NLP) tools and methods. This severely limits the ability of engineers to specify requirements, analyse them, verify compliance, or reason about their interdependencies in a timely and accurate manner without extensive manual efforts.

In this \textbf{innovation opportunity} paper, focusing on the AI-driven innovation opportunities in space RE, we present an approach for small-scale or startup space organisations (like Starbound Space Solutions -- the partner organisation, co-founded by the last author) that often do not have the resources of large-scale space organisations or national agencies.  While these smaller organisations often possess deep domain expertise, they are typically constrained by limited manpower and tool support to manage the end-to-end requirements process. This becomes especially critical when collaborating with larger partners on joint missions. In a typical RE workflow of such projects, the larger organisations (typically known as the \emph{prime} partner) would specify a broader mission scope document (hereafter, $\mathcal{D}$), which lays down the mission details, the key requirements and quality standards from all partners. In such scenarios, smaller organisations (called \emph{subcontractors}) must (1)~identify which parts of the shared mission documents are relevant to their specific subsystems or payloads, (2)~elaborate and adapt these parts to specify the requirements for their parts that reflect their system boundaries and constraints, and (3)~ensure alignment with compliance and documentation standards mandated by the broader mission or governing agencies.

Manually performing these tasks is not only time-consuming but also error-prone, especially when requirements are spread across heterogeneous documents with inconsistent structure and terminology~\cite{abualhaija2020automated}. Small teams often need to sift through hundreds of pages of prime-generated documents, extracting only those requirements that pertain to their payloads or subsystems, interpreting their implications, and translating them into actionable, detailed requirements. The mistakes or omissions in this effort-intensive and error-prone process can lead to integration issues, delays, or even mission failure.

 We focus on this significant yet under-addressed challenge in RE: enabling small or resource-constrained organisations to effectively participate in large, safety-critical engineering projects by understanding, reusing, and aligning with extensive mission documentation. We see a timely opportunity to bridge this gap by integrating recent advances in AI, e.g., \emph{retrieval-augmented generation (RAG)}, \emph{long context embedding}, and \emph{neural labelling} into a modular RE support approach. This approach developed with Starbound (our space industry partner) can also aid other smaller organisations in identifying relevant requirement segments, elaborating on their system-specific needs, and aligning with compliance and quality standards—all with minimal manual overhead.
 
 This paper outlines the core components of this AI-driven approach, demonstrates its feasibility on a real-world document with a preliminary expert evaluation, and presents a roadmap toward a more inclusive, AI-augmented RE workflow for high-assurance domains. Our goal is to initiate a conversation in RE community around novel RE-focused AI toolchains and representations that lower the barrier to entry in increasingly complex systems engineering ecosystems. Our approach's implementation is publicly available\footnote{\url{https://github.com/fanyuuwang/RAGSTAR}}. 

\sectopic{Structure.} Section~\ref{sec:background} provides the background of key technologies underlying our approach. Section~\ref{sec:approach} describes our QA approach. Section~\ref{sec:case_demo} presents a case demonstration and preliminary evaluation discussion, and Section~\ref{sec:roadmap} presents our main roadmap steps and open RE research questions.

\section{Background}~\label{sec:background}
This section covers the background on the relevant NLP concepts used in our approach.
\subsection{Retrieval-Augmented Generation}
Retrieval-Augmented Generation (RAG) has emerged as a powerful technique that combines the knowledge retrieval capabilities of embedding-based search with the generative strengths of large language models (LLMs). Instead of relying solely on an LLM’s pre-training corpus, RAG pipelines retrieve relevant external documents based on a user query and provide them as context for answer generation. This architecture has been successfully applied in different software engineering and RE tasks~\cite{arora2024generating,yang2025ragva}.

\subsection{In-Context Classification}~\label{subsec:ICL}
In-context Learning (ICL), which forms the foundation for in-context classification, has emerged as a powerful paradigm in the era of LLMs~\cite{edwards2024language}. By conditioning on a few demonstration examples or class definitions within the input context, ICL enables LLMs to perform a wide range of tasks—such as classification—without any gradient updates or parameter modifications. This approach is often called in-context classification~\cite{edwards2024language}. The in-context classification task is typically divided into two stages: i) category definition and ii) category prediction~\cite{zhu2023icxml}. In the space industry, the application scenarios and corresponding classes might vary depending on the task. For instance, a document may describe requirements related to payload design, onboard data handling, or autonomy—each representing distinct classification categories. Since these categories are often predefined for a given project, our approach focuses primarily on the prediction stage. We adopt ICXML~\cite{zhu2023icxml} as our in-context classification framework.

\subsection{Label Distribution Expression}
In traditional classification tasks, each document or requirement is typically assigned a single label (e.g., ``Payload'' or ``Autonomy''), or a set of discrete labels. However, in space system requirements—where a single paragraph may touch on multiple overlapping concerns—this all-or-nothing labelling is too rigid. To address this, we adopt a more flexible representation inspired by \emph{label distribution learning}~\cite{zhao2023imbalanced}. Instead of assigning just one category, we assign a distribution of scores across all predefined categories, capturing how strongly each category is expressed in a given document segment. For example, a paragraph might be 70\% relevant to ``Payload'' and 30\% to ``Platform'' rather than being forced into one or the other.
We refer to this representation as a \textbf{neural label}. It allows us to: (1) Quantify the relationship between a document and multiple application categories; (2) Handle overlap in requirements descriptions; (3) Support fine-grained similarity comparisons between documents. This continuous, multi-dimensional label format better reflects the nuanced, multi-purpose nature of real-world requirements documents in the space industry, especially when identifying relevant content across large-scale mission documents.

\section{Approach}~\label{sec:approach}
\begin{figure}
  \includegraphics[width=0.43\textwidth]{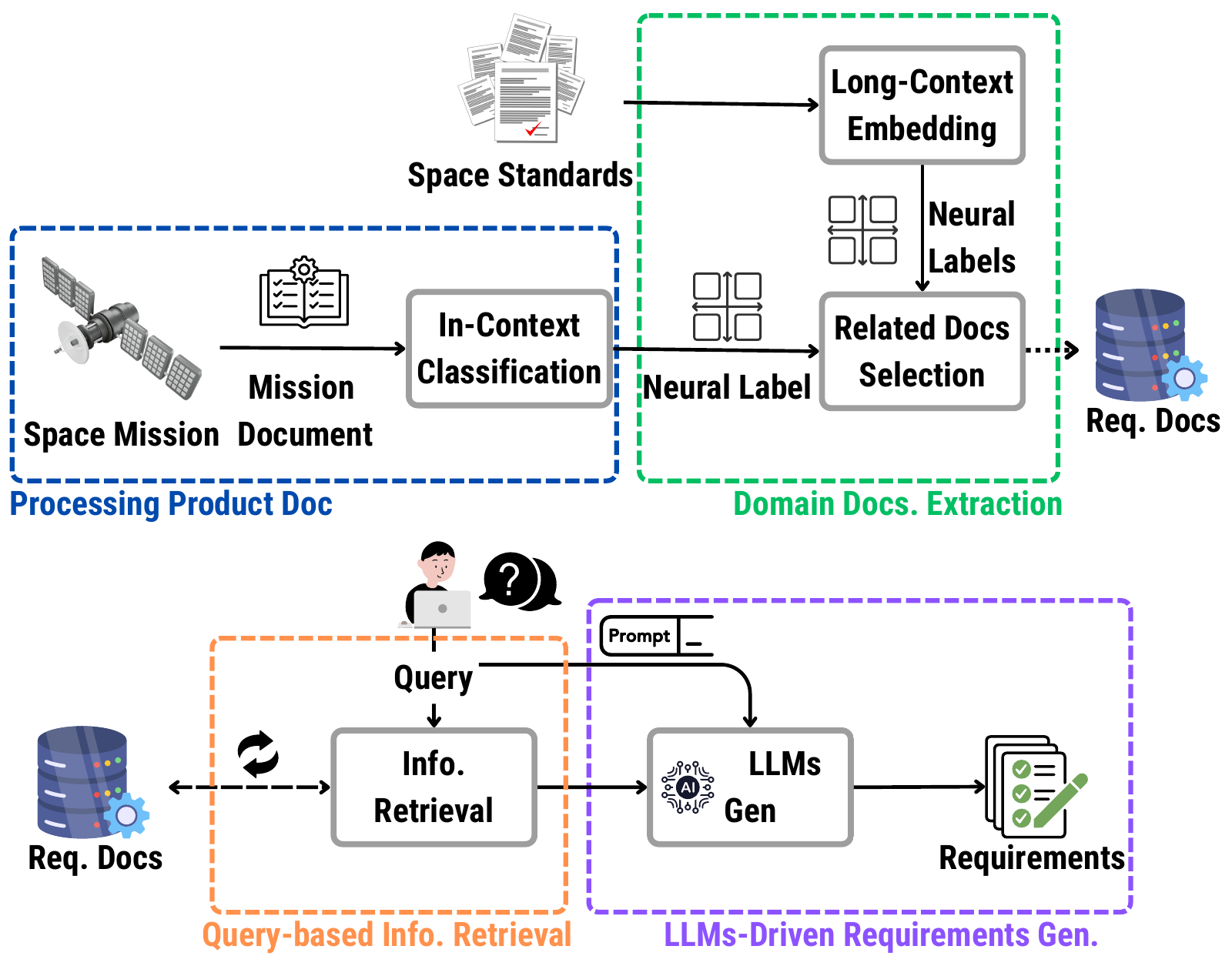}
  \centering
  \vspace*{-.7em}
  \caption{Approach Overview.}
  \vspace*{-1.5em}
  \label{fig:approach}
  \vspace*{-.5em}
\end{figure}

Fig.~\ref{fig:approach} provides an overview of our RAG-based approach for requirements generation. It consists of two pre-processing steps, namely document preprocessing and category consolidation. Both these steps are optional and can be completed semi-automatically. In case, there is a prime mission document ($\mathcal{D}$) available in a machine readable form and the categories are known, these steps can be skipped. Thereafter, the approach has four core steps. All six steps, including two prior and four main steps are discussed below.

\subsection{Prior 1: Documents Preprocessing}
\label{sec:preprocessing}
This (prior) step focuses on preparing the relevant project document for processing. In industrial settings, most documents are maintained in PDF or other formats optimised for distribution and reading. As a result, conventional parsing tools often struggle to extract clean, structured text—leading to fragmented outputs where paragraphs are split, line breaks are misplaced, or semantic boundaries are lost~\cite{abualhaija2020automated}.

To address this, we employ open-source Python packages such as \texttt{pdfplumber} and \texttt{PyPDF2} to extract raw text from the original documents. However, due to common formatting inconsistencies, the resulting text often requires additional restructuring before it can be meaningfully analysed. We leverage an LLM to process this output, to reconstruct coherent paragraphs that more closely resemble the original document structure. While the process significantly reduces manual effort, we still recommend a light manual review to ensure no critical content is lost or misinterpreted in preprocessing.


\subsection{Prior 2: Categories Consolidation}
\label{sec:cate_conso}
Space industrial requirements are highly complex and multifaceted. Even within functional requirements, multiple segments describe distinct application scenarios. In our approach, these application scenarios are predefined into specific categories, e.g., Payload, Orbit-related and On-board data handling. In the subsequent steps, LLMs analyse the descriptions, enabling them to annotate the documents and requirements with corresponding neural labels.

\subsection{Step 1: Processing Mission Document ($\mathcal{D}$)}
\label{sec:proj_doc_proc}
The mission document ($\mathcal{D}$) provides a detailed description of the space mission and is significantly related to the project. Processing $\mathcal{D}$ is the first and crucial step, as the accuracy of its interpretation not only determines the results of identifying related domain documents but also affects the RAG process.
\begin{enumerate}
    \item Given a mission document $\mathcal{D}$ and a set of predefined application categories (also called as scenarios in the domain) $\mathcal{C}$, where $\mathcal{C} = [C_1, C_2,...,C_7]$(as detailed in Section \ref{sec:cate_conso}).
    
    \item $\mathcal{D}$ is segmented into individual chunks (paragraphs), resulting in $\mathcal{D}\to [d_1, d_2,...,d_n]$, where each paragraph is treated as a distinct statement.
    \item Using in-context learning, for each text chunk $[d_1, d_2, \ldots, d_n]$ the definition of application categories $\mathcal{C}$ are respectively extended as $[d_i | C_1, d_i | C_2,\ldots, d_i | C_7]$, and their correlation is computed (see Section~\ref{subsec:ICL}). Each chunk is assigned the category with the highest correlation score, resulting in labels $[c'_1, c'_2, \ldots, c'_n]$ drawn from the predefined categories.
    \item The neural label of $\mathcal{D}$ is computed based on the distribution of the categories assigned to the individual chunks. The counts of $[c'_1, c'_2, \ldots, c'_n]$ are aggregated according to the predefined scenarios $\mathcal{C} = [C_1, C_2, \ldots, C_7]$, resulting in a vector representation for the mission document $\mathcal{D}$. The dimensionality of this vector is equal to the number of predefined scenarios.
\end{enumerate}

\subsection{Step 2: Related Domain Documents Extraction}
\label{sec:domain_docs_query}
The domain documentation in the space industry consists of massive, long-context documents. For example, ECSS (European Cooperation for Space Standardization) comprises more than 50 documents, significantly complicating querying and comprehension processes. To address this challenge, we adopted a long-context embedding method—GTE~\cite{zhang2024mgte}—which offers superior nuanced modelling capabilities, especially when handling extensive textual input. This step is essentially a filtration step, which helps identify only closely related domain documents. This filtration step is crucial to avoid misclassification and missing out on important information for a given subcontractor's requirement category, given the large corpora of domain documentation. 
The query process:
\begin{enumerate}
    \item The domain documents are converted into text embeddings, denoted as ${E}=[E_1, E_2,..., E_m]$.
    \item We employ the same embedding method to encode the descriptions of the predefined categories, and then compute the similarity between these embeddings and the document embeddings. This process yields the neural labels for each document. Each document's label has the same dimensionality as that of $\mathcal{D}$.
    \item Based on the neural labels of both the mission and domain documents, we extract the top-k domain documents with similar distributions.
    \item The parts of the mission document and the extracted domain document will be structured as requirements-relevant documentation ($\mathcal{R}$) for the subcontractor.
\end{enumerate}

\subsection{Step 3: Query-based Information Retrieval}
\label{sec:information_retrieval}
After constructing the related documents for the current mission document, the user query will be used to filter related information from the narrowed $\mathcal{R}$ for answer generation in the next step.
\begin{enumerate}
    \item Generate a query $\mathcal{Q}$ (for a given category $\mathcal{C}_{\mathcal{Q}}$).
    \item The description $\mathcal{C}_{\mathcal{Q}}$ (predefined in Sec.~\ref{sec:cate_conso}) will extend the query, as $\mathcal{Q} | \mathcal{C}_{\mathcal{Q}}$.
    \item An information retrieval model named ICRALM~\cite{ram2023context} will be applied to extended query $\mathcal{Q} | \mathcal{C}_{\mathcal{Q}}$ to retrieval the related content from requirements documentation $\mathcal{R}$.
\end{enumerate}

\subsection{Step 4: LLMs-Driven Requirements Generation}
\label{sec:answer_generation}
In this step, we construct the retrieved product description and the domain standards into prompt as input for LLM. The LLM can provide an answer to our query based on the provided context information. We construct our prompt in a predefined template, where three types of information will be ingested into the template, including i) scenario description, ii) requirements (Product)
description, and iii) domain standards. The specific prompt template is defined as:

\label{list:prompt}
\begin{lstlisting}[style=prompt,caption={Answer Generation Prompt Template}]
<User>: You are a requirements analyst from a satellite communications company named STARBOUND, participating in {MISSION}. Based on the task description, mission-level requirements, and domain standards provided, please identify and summarise all information relevant to the following areas and provide requirements-related information for each section.
<Scenario Description>: {INPUT SCENARIO}
<Requirements>: {INPUT REQUIREMENTS}
<Domain Standards>: {INPUT DOMAIN STANDARDS}
1. Read the provided Scenario Description carefully and understand the task.
2. Identify related content from the provided Requirements and Domain Stand.
3. Structure the response in sections. We want you to generate the requirements-related information in each section. 
\end{lstlisting}
\vspace*{-1em}


\section{Case Demonstration}
\label{sec:case_demo}
In this section, we present a case demonstration using a real space document for a space rideshare program\footnote{https://shorturl.at/A2Y2G}—wherein smaller satellites can be launched alongside primary payloads. In such programs, smaller satellite providers must carefully examine the rideshare integration guidelines, identify relevant requirements applicable to their payload, and ensure compliance with mission-specific and standard guidelines. We use the ECSS documents\footnote{\url{https://ecss.nl/}}, with more than 50 documents, for standardising space projects' documentation. We use this as an example case with a lightweight expert qualitative review to show how our approach assists in using documentation for categorising requirements based on predefined scenarios and retrieving contextually relevant information to support efficient requirements generation. 

\sectopic{Preprocessing.} As specified in Sections~\ref{sec:preprocessing} and \ref{sec:cate_conso}, the rideshare payload user guide and the ECSS documents are converted to plain text.

\sectopic{Step 1.} Based on our in-context classification method, the text chunks in the rideshare user guide are assigned predefined categories (see Section~\ref{sec:cate_conso}), for example, $[1, 1, 2, 2, 0, 1, 2, \dots]$, where each number corresponds to a specific category from these categories: Payload, Platform, Launch Vehicle, Orbit-Related Aspects, On-Board Data Handling, Reference Operation Scenarios / Observation Characteristics, and Operability / Autonomy Requirements. For example, by applying the descriptions of ``Launch Vehicle: The rocket or launch system that delivers the spacecraft into its intended orbit...'' and ``Payload: The mission-specific instruments onboard the spacecraft directly achieve its primary objectives...'' on the text chunk, ``The Launch Vehicle uses a right-hand X-Y-Z coordinate frame...'', the in-context classification method will return the correlation of the text chunk conditioned on two categories. Then, the category ``Launch Vehicle'' (higher score) is selected. We then count the occurrences of each category to obtain the neural label for the user guide, such as $[60, 291, 72, 8, 31, 25, 0]$. This indicates that, according to the LLM’s interpretation, the document primarily discusses Platform, followed by Launch Vehicle, Payload, and other topics. An expert review of top-ranked categories confirmed that the distribution aligns with the document's actual thematic focus. 

\sectopic{Step 2.} Utilising an embedding model designed for long-context processing, the ECSS documents are indexed with text embeddings. We compute the dot product between each document and the embeddings of the predefined categories, resulting in neural labels such as $[-15.7, -19.6, -15.3, -17.0, -11.6, -0.7, -16.9]$. Subsequently, we calculate the cosine similarity between the neural label of the user guide and those of the ECSS documents, selecting only documents with a similarity $>0$.  
These documents, together with the rideshare document, form the basis of the subcontractor's requirements document. For instance, at this stage, this resulted in 17 documents that were relevant to our predefined categories, which form the basis of our RAG retrieval process. A manual review of the top-ranked ECSS documents revealed that the majority contained content semantically aligned with the user guide’s dominant categories. This suggests that the neural labels provide a promising mechanism for identifying contextually relevant domain standards. The identified set did not miss out on any relevant document that an expert would have chosen manually, but it did retrieve documents that were deemed only marginally related.


\sectopic{Step 3.} Given a query regarding payload — covering key concepts of \textit{Payload Design Constraints: Outline requirements related to materials, contamination, vibration, shock, and natural frequency. Include rules regarding pressure vessels, solid propulsion, and safety} — we employ an in-context RAG method to retrieve the most relevant text chunks from both the ShareRide user guide and the ECSS documents within the constructed requirements documentation. Specifically, we retrieve the top-10 and top-20 relevant chunks for the ShareRide user guide and the ECSS documents, respectively. For example:
(1) Certification data for Payload hazardous systems. (2) Payloads must have no elastic natural frequencies below 40 Hz and must have a quality factor (...). (3) Environmental constraints, including the operating environment. (e.g., drop, shock, vibration, ...).
A qualitative review confirmed that the retrieved passages captured key aspects of the query—such as natural frequency and environmental limits—indicating that scenario-extended queries can effectively surface semantically relevant content. A systematic analysis, although difficult for such large documents, is required.



\sectopic{Step 4.} Based on the retrieved content, we construct a prompt using the template provided in List.\ref{list:prompt}. We employed two LLMs for answer generation process, namely OpenAI o1~\cite{jaech2024openai} and Deepseek-R1~\cite{guo2025deepseek}. Both models generated structured responses that referenced the retrieved content and reflected the intent of the payload query. Preliminary analysis showed that the OpenAI o1 output used language and formatting consistent with RE documentation practices, suggesting the potential for downstream usability by analysts in early-stage requirement elaboration. The approach would save substantial effort for a requirements analyst rather than doing it manually from scratch.

\section{Roadmap and Open Questions}~\label{sec:roadmap}
Our approach demonstrates a promising first step toward AI-assisted RE for small-scale organisations participating in complex space missions. While the preliminary demonstration shows feasibility in categorising and generating over real-world documentation, several research and engineering milestones remain. We outline our roadmap and open questions:


\sectopic{RE for the Underserved: Enabling Small Actors}
We aim to develop RE support systems to serve small and resource-constrained teams that cannot afford traditional heavyweight RE practices. 
\textit{Open questions:} How can RE processes be scaled down while maintaining assurance? What is ``just enough'' RE in safety-critical but budget-constrained settings?

\sectopic{Evaluation Without Ground Truth: Rethinking RE Validation in Complex Domains}
We aim to explore new strategies for evaluating RE techniques in the space domain where ground truth is unavailable, incomplete, or proprietary. Expert assessment in such domains is expensive, given limited time and hourly rates. We aim to work on automated approaches with LLMs as judges and lightweight qualitative feedback loops as substitutes for traditional precision/recall benchmarks.
\textit{Open questions:} How can we define ``success'' for AI-supported RE in domains with no gold-standard requirements sets? Can we crowdsource or co-create requirements datasets with domain partners while preserving confidentiality? 

\sectopic{Assuring Trust and Validity in AI-Generated Requirements.} One of the key issues in our research project is to ensure that AI-generated or -augmented requirements are not only helpful but also verifiable, auditable, and compliant with regulatory expectations in high-assurance domains, such as space. We aim to augment our approach for traceable generation, where each AI-generated requirement includes links to the source documents and rationale~\cite{zhou2024trustworthiness,wang2025requirementsdrivenautomatedsoftwaretesting}.
 \textit{Open Questions}: What forms of evidence (e.g., LLM explanations, coverage metrics) are needed to improve trust in AI-augmented RE pipelines?

\sectopic{Generalisation Across Domains}
While our work targets the space sector, the core techniques—document parsing, neural labelling, semantic retrieval, and LLM generation—are domain-agnostic. We plan to adapt our pipeline to other such domains where similar documentation and compliance challenges exist.


\vspace*{-.5em}
\bibliographystyle{IEEEtran}

\bibliography{paper}





%

\end{document}